\documentclass[graybox]{svmult}

\usepackage{mathptmx}       
\usepackage{helvet}         
\usepackage{courier}        
\usepackage{makeidx}         
\usepackage{graphicx}        
\usepackage{multicol}        
\usepackage[bottom]{footmisc}
\usepackage{cite}

\makeindex             

\def\br{\begin{eqnarray}}
\def\er{\end{eqnarray}}

\def\({\left(}
\def\){\right)}
\def\[{\left[}
\def\]{\right]}

\def\bpsi{\bar{\psi}}

\def\l{\lambda}

\def\pa{\partial}

\def\s{\sigma}

\def\sech{\mathrm{sech}}
\def\t{\tau}
\def\th{\theta}

\def\tf{\tilde{f}}

\def\bp{{\bar \p}}

\def\non{\nonumber}
\def\ba{\begin{align}}
\def\ea{\end{align}}
\def\be{\begin{eqnarray}}
\def\ee{\end{eqnarray}}

\def\s{\sigma}
\def\bp{\bar{\psi}}
\def\l{\lambda}

\def\ph{\phi}

\def\pp{\partial}

\begin{document}

\title*{An alternative construction for the Type-II defect matrix for the sshG}

\author{A.R. Aguirre, J.F. Gomes, A.L. Retore, N.I. Spano, and A.H. Zimerman }
\institute{A.R. Aguirre \at Instituto de F\'isica e Qu\'imica, Universidade Federal de Itajub\'a - IFQ/UNIFEI, Av. BPS 1303,  37500-903, Itajub\'a, MG, Brasil. \email{alexis.roaaguirre@unifei.edu.br}
\and J.F. Gomes, A.L. Retore, N.I. Spano, and A.H. Zimerman  \at Instituto de F\'isica Te\'orica - IFT/UNESP, Rua Dr. Bento Teobaldo Ferraz 271, Bloco II, 01140-070, S\~ao Paulo, Brasil. \email{jfg@ift.unesp.br, retore@ift.unesp.br, natyspano@unesp.br, zimerman@ift.unesp.br}}

\maketitle

\abstract*{In this paper we construct a Type-II defect (super) matrix for the supersymmetric sinh-Gordon model as a product of two Type-I defect (super) matrices. We also show that the resulting defect matrix corresponds to a fused defect.}

\abstract{In this paper we construct a Type-II defect (super) matrix for the supersymmetric sinh-Gordon model as a product of two Type-I defect (super) matrices. We also show that the resulting defect matrix corresponds to a fused defect.}

\section{Introduction}
\label{intro}

Integrable classical field theories with defects and its connection with 
Type-I and Type-II B\"acklund transformations (BT) has been widely studied in recent years by using mainly the Lagrangian formalism and the defect matrix approach  \cite{Bow1}--\cite{retore}. The classical  integrability  is ensured by the derivation of modified higher order conserved quantities, which requires explicit solutions for the corresponding defect matrices.

On the other hand, the supersymmetric extensions for Liouville and sinh-Gordon (sshG) models with Type-I and Type-II defects has been also discussed in \cite{Ale6}--\cite{Nathaly2}, and their associated defect matrices constructed.

More recently, it has been proposed in \cite{retoreKdV} that Type-II defect matrices could be  constructed as a product of two Type-I defect matrices. This proposal was checked for the bosonic case of the mKdV hierarchy. 

The aim of this paper is to verify this proposal for the sshG model and show that the resulting defect matrix corresponds to a fused defect.

\section{Type-I  and Type II defect formulation}
The Lagrangian density describing the $N = 1$ sshG model with Type-I defects located at $x=x_1$  can be written
as follows, 
\begin{equation}
 {\mathcal L} = \theta(x_1-x)  {\mathcal L}_1  +\delta(x-x_1){\mathcal L}_{D_1} + \theta(x-x_1){\cal L}_0,
\end{equation}
with
\begin{eqnarray}
 {\mathcal L}_p &=&\frac{1}{2}(\partial_x \phi_p)^2 - \frac{1}{2}(\partial_t \phi_p)^2 +   i\psi_p(\partial_x +\partial_t)\psi_p - i\bar{\psi}_p(\partial_x -\partial_t)\bar{\psi}_p \non \\ && \!\! + 4\left[\cosh(2\phi_p)-1\] -8i\bpsi_p\psi_p\cosh\phi_p,\qquad\mbox{}\label{Lag1}\\
 {\mathcal L}_{D_1} &=&\frac{1}{2}(\phi_0\partial_t\phi_1-\phi_1\partial_t\phi_0) -i\psi_1\psi_0 -i\bar{\psi}_1\bar{\psi}_0 + 2i g_1\pa_t g_1 + B_0^{(1)} +B_1^{(1)}, \mbox{}\label{Lagdef1}
\end{eqnarray}
where $\phi_p$ is a real scalar field, and $\psi_p$, $\bpsi_p$ are the components of a Majorana spinor field in
the regions $x> x_1$ ($p = 0$) and $x <x_1 $ ($p =1$) respectively, and $g_1$ an auxiliary fermionic field defined at the defect point. The defect potentials are given by,
\begin{eqnarray}
 B_0^{(1)}\! &=&\!2\s_1 \cosh(\phi_0+\phi_1)+\frac{2}{\s_1}\cosh(\phi_0-\phi_1) , \\
 B_1^{(1)}\! &=&\! 2i\sqrt{2}g_1\Big[\sqrt{\s_1}\cosh\Big(\frac{\phi_0+\phi_1}{2}\Big)(\bar{\psi}_0+\bar{\psi}_1)+\frac{1}{\sqrt{\s_1}}\cosh\Big(\frac{\phi_0-\phi_1}{2}\Big) (\psi_0-\psi_1)\Big].\non
\end{eqnarray}
where $\s_1$ represent the B\"acklund parameter. Besides the bulk field equations, we get the following defect equations at $x=x_1$,
\begin{eqnarray}
\partial_{t}\phi_{0}-\partial_{x}\phi_{1} &=&2\s_1\sinh(\phi_{0}+\phi_{1})-\frac{2}{\s_1}\sinh(\phi_{0}-\phi_{1})\label{defeito1}\\
&+&\!\!\sqrt{2\s_1}ig_1\!\Big[\!\sinh\!\Big(\frac{\phi_{0}+\phi_{1}}{2}\Big)\!(\bar{\psi}_{0}+\bar{\psi}_{1})-\frac{1}{\s_1}\!\sinh\!\left(\frac{\phi_{0}-\phi_{1}}{2}\right)\!(\psi_{0}-\psi_{1})\Big],\non \\
\partial_{x}\phi_{0}-\partial_{t}\phi_{1} \!&=&\!2\s_1\sinh(\phi_{0}+\phi_{1})+\frac{2}{\s_1}\sinh(\phi_{0}-\phi_{1})\label{defeito2}\\
&+&\!\!\sqrt{2\s_1}ig_1\!\Big[\!\sinh\!\Big(\frac{\phi_{0}+\phi_{1}}{2}\Big)\!(\bar{\psi}_{0}+\bar{\psi}_{1})+\frac{1}{\s_1}\!\sinh\!\left(\frac{\phi_{0}-\phi_{1}}{2}\right)\!(\psi_{0}-\psi_{1})\Big],\non \\
\psi_{0}+\psi_{1} & = & 2\sqrt{\frac{2}{\s_1}}\,\cosh\Big(\frac{\phi_{0}-\phi_{1}}{2}\Big)g_{1},
\label{defeito3}\\
\bar{\psi}_{0}-\bar{\psi}_{1} & = & -2\sqrt{2\s_1}\,\cosh\Big(\frac{\phi_{0}+\phi_{1}}{2}\Big)g_{1},
\label{defeito4}\\
\partial_{t}g_{1}\! &=&\!\sqrt{\frac{\s_1}{2}}\Big[\frac{1}{{\s_1}}\cosh\!\Big(\frac{\phi_0-\phi_1}{2}\Big)\! (\psi_1-\psi_0)-\cosh\!\Big(\frac{\phi_0+\phi_1}{2}\Big)\!(\bar{\psi}_0+\bar{\psi}_1)\Big]\!.\mbox{}
\label{defeito5} 
\end{eqnarray}
These defect conditions preserve the integrability of the system after considering defect contributions to the conserved quantities \cite{Nathaly1}. The generating function for an infinite set of
modified conserved quantities depends on the existence of the defect matrix ${K}_1$ connecting two field configurations, namely $\Psi^{(0)}= K_1\Psi^{(1)}$, satisfying the following equations,
\br
 \pa_\pm K_1 = K_1 A_\pm^{(1)} - A_\pm^{(0)}K_1, 
\er
where $\pa_\pm =\frac{1}{2}(\pa_x\pm\pa_t)$, $\l$ is a spectral parameter, and $\Psi^{(p)}$ are vector-valued fields satisfying the associated auxiliary linear problem, $\pa_\pm \Psi^{(p)}=- A_\pm^{(p)} \Psi^{(p)}$. The Lax pair $A_\pm^{(p)}$ are $3\times 3$ graded matrices valued in the $sl(2, 1)$ Lie superalgebra, which can be written in the following form,
\br
{A}_{+}^{(p)} &=&\left(\begin{array}{cc|c}\lambda^{1/2}-\pp_{+}\phi_p&-1&\sqrt{i}\bp_p\\[0.2cm] -\lambda&\lambda^{1/2}+\pp_{+}\phi_p&\l^{1/2}\sqrt{i}\bp_p \tabularnewline\hline \mbox{}&\mbox{}&\mbox{}\\[-0.3cm]
	\l^{1/2}\sqrt{i}\bp_p&\sqrt{i}\bp_p& 2\lambda^{1/2}       \end{array}\right)\label{lax +}, \\[0.3cm]
{A}_{-}^{(p)}&=&\left(\begin{array}{cc|c}\lambda^{-1/2}& -\l^{-1} e^{2 \phi_p}&    \l^{-1/2}\sqrt{i} \psi_p \,e^{\phi_p}\\[0.2cm]- e^{-2 \phi_p} & \l^{-1/2}& \sqrt{i} \psi_p  \,e^{-\phi_p}  \tabularnewline\hline \mbox{}&\mbox{}&\mbox{}\\[-0.3cm]
	-\sqrt{i}  \psi_p \,e^{-\phi_p}  & -\sqrt{i} \l^{-1/2}\psi_p\,e^{\phi_p}& 2\l^{-1/2}\end{array}\right)\label{lax -}.
\er
Therefore, we find that a suitable solution for the type-I defect matrix K can be
written in the following explicit form \cite{Nathaly1},
\begin{eqnarray}
K_1 =c_1\l^{1/2}\left(\begin{array}{cc|c}
 1& \frac{\s_1}{\l} e^{\phi_1+\phi_0} & -\sqrt{\frac{2i\s_1}{\l}} e^{\frac{\phi_1+\phi_0}{2}}g_1\\[0.3cm]
 \s_1\,e^{-(\phi_1+\phi_0)}  & 1 &  -\sqrt{2i\s_1}e^{-\frac{(\phi_1+\phi_0)}{2}}g_1\tabularnewline\hline \mbox{}&\mbox{}&\mbox{}\\[-0.2cm]
\sqrt{2i\s_1} e^{-\frac{(\phi_1+\phi_0)}{2}}g_1\quad \mbox{}& \sqrt{\frac{2i\s_1}{\l}}e^{\frac{(\phi_1+\phi_0)}{2}}g_1 & 1-\frac{\s_1}{\l^{1/2}} \\
\end{array} \right),\quad\mbox{}\label{1.3}
\end{eqnarray}
where $c_1$ is a free constant parameter.

Now, the Type-II defect for the $N=1$ sshG model can be constructed by considering initially a two-defects system of the Type-I at different points, and then fusing them to the same point by taking a limit in the Lagrangian density \cite{Nathaly2}--\cite{Robertson}. Let us consider one of the defects placed at $x=x_1$ and the other at $x=x_2$. The Lagrangian density for this system can be written as follows,
\begin{eqnarray}
 {\mathcal L} &=& \theta(x_1-x)  {\mathcal L}_1  +\delta(x-x_1){\mathcal L}_{D_1} + \theta(x-x_1)\th(x_2-x){\cal L}_0 \non \\ &&+\delta(x-x_2) {\cal L}_{D_2}+ \theta(x-x_2) {\mathcal L}_2, 
\end{eqnarray}
where ${\cal L}_p$, with $p=0,1,2$, is given by eq. (\ref{Lag1}), and the two type-I defect Lagrangian densities at $x=x_k$, $k=1,2$, are given by eq. (\ref{Lagdef1}). Now, we have two auxiliary fermionic fields $g_k$, and two free parameters $\s_k$, with $k=1,2$, defined at the defect positions, respectively. At Lagrangian level, the fusing of defects can be performed by taking the limit \mbox{$x_2\to x_1$}. After some manipulations, it was shown that the fused defect is equivalent to a type-II defect \cite{Nathaly2}, and takes the following form
\begin{eqnarray}
 {\cal L}_{D} &=&  \!\phi_- \pa_t \l_0 -\frac{1}{2}\phi_-\pa_t \phi_+ + \frac{i}{2}(\bar{\psi}_+\bar{\psi}_- - \psi_+\psi_-) + if_1\pa_t f_1 +i\tf_1\pa_t\tf_1 +B,
 \label{f4.39}
\end{eqnarray}
with $\phi_\pm=\phi_1\pm\ph_2$, $\psi_\pm=\psi_1\pm\psi_2$, and  $B = B_0^{(+)} + B_0^{(-)}  + B_1^{(+)}+B_1^{(-)}$the defect potentials,
\begin{eqnarray}
 B_0^{(+)}&=&  {m\s}\left[e^{(\phi_+-\l_0)} + e^{-(\phi_+-\l_0)}\Big(\sinh^2\Big(\frac{\phi_-}{2}\Big) +\cosh^2\tau \Big)\right],\label{f4.40}\\
B_0^{(-)} &=& \frac{m}{\s}\left[e^{-\l_0} + e^{\l_0}\Big(\sinh^2\Big(\frac{\phi_-}{2}\Big) +\cosh^2\tau \Big)\right],\\[0.1cm]
 B_1^{(+)} &=& -i\sqrt{m\s}\left[\Big(e^{\frac{\left(\phi_+-\l_0\)}{2}}+e^{-\frac{\left(\phi_+-\l_0\)}{2}}\cosh\tau\Big)\bpsi_+f_1 +e^{-\frac{\left(\phi_+-\l_0\)}{2}}\sinh\Big(\frac{\phi_-}{2}\Big)\bpsi_+\tf_1\right]\non\\
  && + im \s\Big( 1+ {e^{-(\phi_+-\l_0)}} \cosh \tau \Big)\cosh\Big(\frac{\phi_-}{2}\Big)f_1\tf_1, \\[0.1cm]
B_1^{(-)}&= &-i\sqrt{\frac{m}{\s}}\left[\Big(e^{-{\l_0\over 2}}+e^{{\l_0\over 2}}\cosh\t\Big)\psi_+\tf_1-e^{{\l_0\over 2}}\sinh\Big(\frac{\phi_-}{2}\Big)\psi_+f_1\right]\non \\
 && + \frac{ im}{\s}\Big( 1+  {e^{\l_0}}\cosh \tau \Big)\cosh\Big(\frac{\phi_-}{2}\Big)f_1\tf_1,\label{f4.43}
\end{eqnarray}
where it has been used $\s_1 = \s e^{-\tau},  \s_2=\s\,e^{\tau}$, and the reparametrizations
\begin{eqnarray}
 \phi_0 &\to& -\l_0+\frac{\phi_+}{2}-\ln\left[\cosh\Big(\frac{\phi_-}{2}-\tau\Big) \right]- \frac{i}{2}{\rm \sech}\Big(\frac{\phi_-}{2}-\t\Big) f_1\tf_1, \label{f4.22}\\
 f_1 &=& \mu_+ g_2 +\mu_- g_1, \qquad 
 \tf_1 =  \mu_- g_2-\mu_+ g_1, \qquad  \mu_\pm =\left[\frac{1+e^{\pm(\phi_- -2\t)}}{2}\]^{-\frac{1}{2}}.\quad\,\, \mbox{}\label{rep}
\er
From the above defect Lagrangian we can write the defect conditions at $x_1=x_2$,
\begin{eqnarray}
( \pa_x -\pa_t)\phi_+ \!&=&\! \pa_t\l_0 -m\!\left[\s\,e^{-(\phi_+-\l_0)}+\frac{1}{\s}e^{\l_0}\]\!\sinh\phi_-  -im\Big(\s+\frac{1}{\s}\Big)\!\sinh\!\Big(\frac{\phi_-}{2}\Big) f_1\tf_1\non\\
&&  +i\sqrt{m\s}\, e^{-\frac{\left(\phi_+-\l_0\)}{2}}\cosh\Big(\frac{\phi_-}{2}\Big)\bpsi_+\tf_1 -i\sqrt{m\over \s}e^{\l_0 \over 2}\cosh\Big(\frac{\phi_-}{2}\Big)\psi_+f_1\qquad\mbox{} \non\\
&&\!-im\left[\s \,e^{-(\phi_+-\l_0)} +\frac{1}{\s}e^{\l_0}\right]\cosh\t\sinh\Big(\frac{\phi_-}{2}\Big) f_1\tf_1,\qquad \mbox{}\label{f4.45}\\
 (\pa_x+\pa_t)\phi_- &=& 2m\s\left[ e^{-(\phi_+-\l_0)}\Big(\sinh^2\Big(\frac{\phi_-}{2}\Big) +\cosh^2\tau \Big)-e^{(\phi_+-\l_0)} \right]\non\\
 && +i\sqrt{m\s}\Big(e^{\frac{\left(\phi_+-\l_0\)}{2}}-e^{-\frac{\left(\phi_+-\l_0\)}{2}}\cosh\tau\Big)\bpsi_+f_1 \non \\
 && -i\sqrt{m\s}\,e^{-\frac{\left(\phi_+-\l_0\)}{2}}\sinh\Big(\frac{\phi_-}{2}\Big)\bpsi_+\tf_1\non\\
 && +2im\s \,e^{-(\phi_+-\l_0)}\cosh\t \cosh\Big(\frac{\phi_-}{2}\Big)f_1\tf_1,\\[0.1cm]
 (\pa_x-\pa_t)\phi_-   &=& \frac{2m}{\s}\left[e^{-\l_0} - e^{\l_0}\Big(\sinh^2\Big(\frac{\phi_-}{2}\Big) +\cosh^2\tau \Big)\right]\non\\
 && -i\sqrt{\frac{m}{\s}}\left[\Big(e^{-{\l_0\over 2}}-e^{{\l_0\over 2}}\cosh\t\Big)\psi_+\tf_1 +e^{{\l_0\over 2}}\sinh\Big(\frac{\phi_-}{2}\Big)\psi_+f_1\right]\non\\
 && -\frac{2im}{\s }\,e^{\l_0}\cosh\t \cosh\Big(\frac{\phi_-}{2}\Big)f_1\tf_1,\\[0.1cm]
 \psi_-  &=& \sqrt{\frac{m}{\s}}\left[e^{{\l_0\over 2}}\sinh\Big(\frac{\phi_-}{2}\Big)f_1-\Big(e^{-{\l_0\over 2}}+e^{{\l_0\over 2}}\cosh\t\Big)\tf_1\right],\\[0.1cm]
 \bpsi_- &=&\!\!\sqrt{m\s}\!\left[\!\Big(e^{\frac{\left(\phi_+-\l_0\)}{2}}\!+e^{-\frac{\left(\phi_+-\l_0\)}{2}}\cosh\tau\Big)f_1 +e^{-\frac{\left(\phi_+-\l_0\)}{2}}\sinh\Big(\frac{\phi_-}{2}\Big)\!\tf_1\right]\!\!,\quad \mbox{}\\[0.1cm]
 \pa_t f_1 &=& -\frac{\sqrt{m\s}}{2}\Big(e^{\frac{\left(\phi_+-\l_0\)}{2}}+e^{-\frac{\left(\phi_+-\l_0\)}{2}}\cosh\tau\Big)\bpsi_+ +\frac{1}{2}\sqrt{\frac{m}{\s}}
 e^{{\l_0\over 2}}\sinh\Big(\frac{\phi_-}{2}\Big)\psi_+\non \\
 &&-\frac{m}{2} \left[\Big( \s+\frac{1}{\s}\Big) + \Big(\s {e^{-(\phi_+-\l_0)}} +\frac{1}{\s}  {e^{\l_0}}\Big)\cosh \tau\right]\!\cosh\Big(\frac{\phi_-}{2}\Big)\tf_1, \\[0.1cm]
 \pa_t\tf_1   &=&  -\frac{\sqrt{m\s}}{2}e^{-\frac{\left(\phi_+-\l_0\)}{2}}\sinh\Big(\frac{\phi_-}{2}\Big)\bpsi_+ -\frac{1}{2}\sqrt{\frac{m}{\s}}\Big(e^{-{\l_0\over 2}}+e^{{\l_0\over 2}}\cosh\t\Big)\psi_+\non \\
 &&  +\frac{m}{2}  \left[\Big( \s+\frac{1}{\s}\Big) + \Big(\s {e^{-(\phi_+-\l_0)}} +\frac{1}{\s}  {e^{\l_0}}\Big)\cosh \tau\right]\!\cosh\Big(\frac{\phi_-}{2}\Big)f_1.\label{f4.51}
\end{eqnarray}
%
In order to derive the associated Type-II defect super-matrix for the model,  we propose \cite{retoreKdV} to construct it as a product of two Type-I  defect matrices, such that
\br
  \Psi^{(2)} &=& K_1(\s_2) \Psi^{(0)} \, = \, K_1(\s_2) K_1(\s_1) \Psi^{(1)} \,= \, K_2(\s,\t)\Psi^{(1)},
\er 
where $K_2(\s,\t)= K_1(\s_2) K_1(\s_1)$. Therefore, by a direct computation we find that the components $k_{ij}$ of the fused defect matrix $K_2$ are given by,
\br 
            k_{11} &=&c \left(\l +  \s^2 e^{-\phi_-} + 2 i \s \,e^{-\frac{\phi_-}{2}} (g_1g_2) \l^{1/2}\),\\
            k_{12}&=&c\s\, e^{\phi_0}\left(e^{(\phi_1-\t)} + e^{(\phi_2+\t)} + 2i\, e^{\frac{\phi_+}{2}}\,(g_1g_2)\),\qquad \mbox{}\\
             k_{13} &=& -c\s\sqrt{2i\s}\,e^{\phi_0 \over 2}\Big( \, e^{\phi_2-\frac{(\phi_1-\t)}{2}}g_1-\,e^{\frac{(\phi_2-\t)}{2}}g_2\Big)\\&&
              - c\sqrt{2i\s} \l^{1/2}\,e^{\phi_0 \over 2}\Big( e^{\frac{(\phi_1-\t)}{2}}g_1 +e^{\frac{(\phi_2+\t)}{2}}g_2\Big)\\
               k_{21}&=&c\s\,e^{-\phi_0}\left(e^{-(\phi_1+\t)}+ e^{-(\phi_2-\t)}+2i\, e^{-\frac{\phi_+}{2}}g_1g_2\),\\
                k_{22}&=&c\left(\l +\s^2 e^{\phi_-}+2i\s \,e^{-\frac{\phi_-}{2}}g_1g_2\), \\
                 k_{23}&=&-c\sqrt{2i\s}\,\l \,e^{-\frac{\phi_0}{2}}\Big(\,g_1 e^{-\frac{(\phi_1+\t)}{2}} + \,g_2 e^{-\frac{(\phi_2-\t)}{2}}\Big)\nonumber\\&&+
			 c\s\sqrt{2i\s}\,\l^{1/2}\, e^{-\frac{\phi_0}{2}}\Big(\,g_2 e^{-\frac{(\phi_2+\t)}{2}} -\, g_1 e^{\frac{(\phi_1+\t)}{2}-\phi_2} \Big),\\
			  k_{31}&=&c\sqrt{2i\s}\,\l \,e^{-\frac{\phi_0}{2}}\Big(g_1 e^{-\frac{(\phi_1+\t)}{2}} +g_2 e^{-\frac{(\phi_2-\t)}{2}}\Big)\nonumber\\&&+
			 c\s\sqrt{2i\s}\,\l^{1/2}\,e^{-\frac{\phi_0}{2}}\Big( g_2 \,e^{\frac{(\phi_2-\t)}{2}-\phi_1} - g_1 \,e^{\frac{(\phi_1+\t)}{2}} \Big),\\
			  k_{32}&=& c\s\sqrt{2i\s}\,e^{\frac{\phi_0}{2}}\Big(g_2\,e^{-\frac{(\phi_2+\t)}{2}+\phi_1}-g_1\,e^{\frac{(\phi_1+\t)}{2}}\Big)\nonumber\\&&+
			 c\sqrt{2i\s}\,\l^{1/2}\,e^{\frac{\phi_0}{2}}\Big(e^{\frac{(\phi_1-\t)}{2}}g_1 +e^{\frac{(\phi_2+\t)}{2}}g_2\Big),\\
			 k_{33}&=& c\(\l+\s^2-2\s \l^{1/2}\Big(\cosh(\t)-2i g_1g_2\cosh\Big(\frac{\phi_-}{2}\Big)\),
          \er
where $c=c_1c_2$. By straightforward comparison with eq. (A.80)--(A.89) in \cite{Nathaly2}, it is not difficult to see that the fused defect matrix derived as product of two type-I defect matrices is equivalent (up to $\l^{1/2}$) to the type-II defect matrix previously found in \cite{Nathaly2}, after reparametrazing the auxiliary fields given as in eqs. \mbox{(\ref{f4.22}) and (\ref{rep}).}


\begin{acknowledgement}
The authors would like to thank the organizers of the colloquium ICGTMP - Group 31 for the opportunity to present our work. ALR would like to thank FAPESP for the financial support under the process 2015/00025-9. JFG would like to thank FAPESP and CNPq for the financial support. NIS and AHZ would like to thank CNPq for the financial support.

\end{acknowledgement}
\end{document}